\documentstyle[12pt,a41,axodraw]{article}

\begin{document}

\thispagestyle{empty}

\begin{flushleft}
YITP-SB-02-58 \hfill
\\INLO-PUB-8/02 \hfill
{\tt hep-ph/0210369}\\
October 2002
\end{flushleft}

\setcounter{page}{0}

\mbox{}
\vspace*{\fill}
\begin{center}
{\LARGE\bf NLO corrections to differential cross sections}

\vspace{2mm}
{\LARGE\bf for pseudo-scalar Higgs boson production}

\vspace*{20mm}
\large
{B. Field , J. Smith,  M.E. Tejeda-Yeomans \footnote {partially supported by 
the National Science Foundation grant PHY-0098527}} 
\\
\vspace{2em}
\normalsize
{\it C.N. Yang Institute for Theoretical Physics}\\
{\it State University of New York at Stony Brook, New York 11794-3840, USA}
\\
\vspace*{20mm}
\large
{W.L. van Neerven }
\\
\vspace{2em}
\normalsize
{\it Instituut-Lorentz}\\
{\it University of Leiden, P.O. Box 9506, 2300 RA Leiden, The Netherlands}

\vspace*{\fill}
\end{center}
\begin{abstract}
\noindent
We have computed the full next-to-leading (NLO) QCD corrections to the 
differential distributions $d^2\sigma/(dp_T~dy)$ for pseudo-scalar 
Higgs (A) production at large hadron colliders. This calculation has been 
carried out using the effective Lagrangian approach which is valid as 
long as the mass of the pseudo-scalar Higgs boson $m_{\rm A}$ and 
its transverse momentum $p_T$ do not exceed the top-quark mass $m_t$.
The shape of the distributions hardly differ from those obtained for
scalar Higgs (H) production because, apart from the overall coupling
constant and mass, there are only small differences between the partonic 
differential distributions for scalar and pseudo-scalar production. Therefore 
there are only differences in the magnitudes of the hadronic differential 
distributions which can be mainly attributed to the unknown mixing angle 
$\beta$ describing the pseudo-scalar Higgs coupling to the top quarks.
\end{abstract}
\vspace*{\fill}

\newpage\noindent
The scalar Higgs boson ${\rm H}$, which is the corner stone of the standard 
model, is the only particle which has not yet been observed. 
Its discovery or its absence will shed light on the mechanism 
how particles acquire mass as well as 
answer questions about super-symmetric extensions of the standard model or
about the compositeness of the existing particles and the Higgs boson. Among
these two alternatives super-symmetry is the most appealing one, in 
particular the minimal super-symmetric extension of the standard model.
The latter version contains two complex Higgs doublets instead of one
and it is therefore called the Two-Higgs-Doublet Model (2HDM). Here
the scalar particle spectrum contains both the Higgs boson H
and another neutral scalar boson ${\rm h}$. Furthermore it
contains two charged scalar bosons ${\rm H}^{\pm}$ and a neutral 
pseudo-scalar Higgs boson ${\rm A}$. The tree-level masses are expressed 
in two independent parameters, namely the mass $m_{\rm A}$
and the ratio of the vacuum expectation values of the two Higgs doublets 
defined by $\tan \beta =v_2/v_1$ (see e.g. \cite{ghks}).
According to the experiments at LEP their 
parameter ranges are restricted so that $m_{\rm A}<91.9~{\rm GeV/c^2}$ 
and $0.5<\tan \beta < 2.4$ \cite{hag} are excluded.
In this paper we study A-production which in lowest order 
proceeds via gluon-gluon fusion where the gluons are coupled to the A via
a heavy flavour triangular loop. This is similar to H-production
except that now the coupling constant describing the
interaction of the A with the quarks depends on both
the masses of the quarks and on the angle $\beta$. This follows from
the 2HDM where the coupling constants of the up and down quarks behave like
$g_{up}\sim m_u \cot \beta$ and $g_{down}\sim m_d \tan \beta$ respectively
\cite{ghks}. Since the effective Lagrangian approach below is only 
valid in the case the mass of the quark appearing in the triangular 
loop satisfies the condition $m_q \gg m_{\rm A}$, the bottom quark is 
excluded. However then we have to require that in the 2HDM the coupling 
of the A to the top-quark is stronger than to the 
bottom-quark which implies the condition
\begin{eqnarray}
\frac{m_t}{m_b}\gg \tan^2 \beta\,.
\end{eqnarray}
If we choose $m_b=4.5~{\rm GeV/c^2}$ and $m_t=173.4~{\rm GeV/c^2}$ one obtains
the inequality $\tan \beta \ll 6.21$. In view of the experimental boundaries
above one can conclude that the results of the calculation below can be only
applied for the regions $\tan \beta < 0.5$ and $2.4 < \tan \beta < 6.21$.  

In the effective Lagrangian approach scalar H-production is described
by the Lagrangian density \cite{dawson}, \cite{chkn1}
\begin{eqnarray}
{\cal L}^{\rm H}_{eff}=G_{\rm H}\,\Phi^{\rm H}(x)\,O(x)\,, \quad \mbox{with} 
\quad
O(x)=-\frac{1}{4}\,G_{\mu\nu}^a(x)\,G^{a,\mu\nu}(x)\,,
\end{eqnarray}
whereas pseudo-scalar A-production is obtained from \cite{spdj}, \cite{kasc},
\cite{chkn2}
\begin{eqnarray}
&& {\cal L}^{\rm A}_{eff}=\Phi^{\rm A}(x)\,\Big [
G_{\rm A}\,O_1(x)+\tilde G_{\rm A}\,O_2(x)\Big ]\,, \quad \mbox{with} \quad
\nonumber\\[2ex]
&& O_1(x)=-\frac{1}{8}\,\epsilon_{\mu\nu\lambda\sigma}\,
G_a^{\mu\nu}(x)\,G_a^{\lambda\sigma}(x)\,,\qquad O_2(x)=
-\frac{1}{2}\partial^{\mu}
\sum_{i=1}^{n_f}\bar q_i(x)\gamma_{\mu}\gamma_5 q_i(x)\,,
\end{eqnarray}
where $\Phi^{\rm H}(x)$ and $\Phi^{\rm A}(x)$ are the scalar and pseudo-scalar
fields respectively and $n_f$ denotes the number of light flavours.
Up to NLO  the operator $O_2(x)$ only contributes when it interferes
with the operator $O_1(x)$ provided the quarks are massless. 
The effective couplings $G_{\rm B}$ (${\rm B}={\rm H}, {\rm A}$) are 
determined by the top-quark triangular graph describing the decay 
process $B\rightarrow g + g$
in the limit $m_t\rightarrow \infty$
\begin{eqnarray}
G_{\rm B}^2=4\,\sqrt 2\,\left (\frac{\alpha_s(\mu_r^2)}{4\pi}\right )^2\,
G_F\,\tau_{\rm B}^2\,F_{\rm B}^2(\tau_{\rm B})\,{\cal C}_{\rm B}^2
\left (\alpha_s(\mu_r^2),
\frac{\mu_r^2}{m_t^2}\right )\,, 
\qquad \tau_{\rm B}=\frac{4\,m_t^2}{m_{\rm B}^2}\,,
\qquad {\rm B}={\rm H},{\rm A}\,.
\end{eqnarray}
and the functions $F_{\rm B}$ are defined by
\begin{eqnarray}
&& F_{\rm H}(\tau)=1+(1-\tau)\,f(\tau)\,,
\qquad F_{\rm A}(\tau)=f(\tau)\, \cot \beta\,,
\nonumber\\[2ex]
&&f(\tau)=\arcsin^2 \frac{1}{\sqrt\tau}\,, \quad \mbox{for} \quad \tau \ge 1\,,
\nonumber\\[2ex]
&& f(\tau)=-\frac{1}{4}\left ( \ln \frac{1-\sqrt{1-\tau}}{1+\sqrt{1-\tau}}
+\pi\,i\right )^2\,, \quad \mbox{for} \quad \tau < 1\,.
\end{eqnarray}
In the large $m_t$-limit $F(\tau)$ behaves as
\begin{eqnarray}
 \mathop{\mbox{lim}}\limits_{\vphantom{\frac{A}{A}} \tau \rightarrow \infty}
F_{\rm H}(\tau)=\frac{2}{3\,\tau}\,,\qquad
 \mathop{\mbox{lim}}\limits_{\vphantom{\frac{A}{A}} \tau \rightarrow \infty}
F_{\rm A}(\tau)=\frac{1}{\tau}\,\cot \beta\,.
\end{eqnarray}
Here $m$ and $m_t$ denote the masses of the (pseudo-) scalar
Higgs boson and the top quark respectively. The running coupling constant is 
given by $\alpha_s(\mu_r^2)$ where $\mu_r$ denotes the renormalization scale 
and $G_F$ is the Fermi constant. The coefficient functions ${\cal C}_{\rm B}$
originate from the corrections to the top-quark triangular graph provided
one takes the limit $m_t \rightarrow \infty$.
We have presented the couplings $G_{\rm B}$ in Eq. (4) for general $m_t$ on the
Born level only in order to keep some part of the top-quark mass dependence. 
This is an approximation because the gluons which couple to the (pseudo-) scalar
Higgs boson via the top-quark loop in the partonic subprocesses are very often 
virtual. The virtual-gluon momentum dependence is neither described by
$F_{\rm B}(\tau)$ nor by ${\cal C}_{\rm B}$. For on-mass-shell gluons the 
latter quantity has been computed in the large $m_t$  
limit up to order $\alpha_s$
in \cite{dawson}, \cite{spdj}, \cite{kasc} and up to order $\alpha_s^2$ in 
\cite{chkn1}, \cite{chkn2}. For our NLO calculations we only
need these coefficient functions corrected up to order $\alpha_s$ and they read
\begin{eqnarray}
{\cal C}_{\rm H}\left (\alpha_s(\mu_r^2),\frac{\mu_r^2}{m_t^2}\right )&=&
1+\frac{\alpha_s^{(5)}(\mu_r^2)}{4\pi}\,\Big (11\Big )+\cdots\,,
\\[2ex]
{\cal C}_{\rm A}\left (\alpha_s(\mu_r^2),\frac{\mu_r^2}{m_t^2}\right )&=& 1\,,
\end{eqnarray}
where $\alpha_s^{(5)}$ is presented in a five-flavour number scheme. 
Notice that Eq. (8) holds in all orders because of the Adler-Bardeen theorem
\cite{adba}. The effective Lagrangian approach has been successfully applied
to compute the total cross section of scalar Higgs production in hadron-hadron 
collisions in NLO \cite{dawson} and NNLO \cite{ha}, \cite{cafl}, \cite{haki1},
\cite{haki2}, \cite{anme1}.
In the case of pseudo-scalar Higgs production this cross section was computed
in NLO in \cite{spdj}, \cite{kasc} and in NNLO in \cite{haki3}, \cite{anme2}.

In this paper we study the semi-inclusive reaction with one pseudo-scalar
Higgs boson ${\rm A}$ in the final state which is given by
\begin{eqnarray}
{\rm H}_1(p_1)+{\rm H}_2(p_2)\rightarrow {\rm A}(-p_5) + 'X'\,,
\end{eqnarray}
where ${\rm H}_1$ and ${\rm H}_2$ denote the incoming hadrons and $X$ 
represents an inclusive hadronic final state. Further we define the following
kinematical invariants
\begin{eqnarray}
S=(p_1+p_2)^2\,, \qquad T=(p_1+p_5)^2\,, \qquad U=(p_2+p_5)^2\,.
\end{eqnarray}
The latter two invariants can be expressed in terms of the transverse momentum 
$p_T$ and rapidity $y$ variables as
\begin{eqnarray}
&&T=m^2-\sqrt S\,\sqrt{p_T^2+m^2}\,\cosh y+\sqrt S\,\sqrt{p_T^2+m^2}
\,\sinh y\,,
\nonumber\\[2ex]
&&U=m^2-\sqrt S\,\sqrt{p_T^2+m^2}\,\cosh y-\sqrt S\,\sqrt{p_T^2+m^2}
\,\sinh y\,,
\end{eqnarray}
where $m$ is the mass of the pseudo-scalar Higgs boson.
The hadronic cross section is given by
\begin{eqnarray}
&& S^2 \frac{d^2~\sigma^{{\rm H_1H_2}}}{d~T~d~U}(S,T,U,m^2)= 
\nonumber\\[2ex]
&&\sum_{a,b=q,g}
\int_{x_{1,{\rm min}}}^1 \frac{dx_1}{x_1} \int_{x_{2,{\rm min}}}^1
\frac{dx_2}{x_2}\, f_a^{\rm H_1}(x_1,\mu^2)
\, f_b^{\rm H_2}(x_2,\mu^2)\,s^2
\frac{d^2~\sigma_{ab}}{d~t~d~u} (s,t,u,m^2,\mu^2)\,,
\nonumber\\[2ex]
&&\mbox{with}\quad x_{1,{\rm min}}=\frac{-U}{S+T-m^2}\,, \qquad
x_{2,{\rm min}}=\frac{-x_1(T-m^2)-m^2}{x_1S+U-m^2}\,,
\end{eqnarray}
where $s$, $t$ and $u$ are the partonic analogues of $S$, $T$ and $U$ 
in Eq. (10)
where $p_1$ and $p_2$ now represent the incoming parton momenta. Further
$f_a^{\rm H_i}$ denotes the parton density corresponding to hadron ${\rm H_i}$
and $\mu$ stands for the factorization scale which for convenience is
set equal to the renormalization scale $\mu_r$ appearing in Eq. (4).
The NLO corrections to the partonic cross section $d^2\sigma/(dt~du)$ 
in the case of H-production based on the effective Lagrangian 
in Eq. (2) are presented in \cite{flgra} and \cite{rasm}.
Here we will give the corresponding results for the ${\rm A}$ described by the
Lagrangian in Eq. (3). The calculation proceeds in the same way as presented
in \cite{rasm}. We use $n$-dimensional regularization 
in order to compute the loop
and phase space integrals which contain ultraviolet, infrared and collinear
singularities. However there is one extra complication in the 
pseudo-scalar case. This concerns the Levi-Civita tensor in Eq. (3) which
is essentially a four dimensional object. Here we follow the same prescription
as in Eq. (4) in \cite{haki3}, \cite{anme2} where the product of two 
Levi-Civita tensors
is contracted in $n$ dimensions if one sums over dummy Lorentz indices. This
prescription leads to an interference between diagrams carrying the vertex
coming from the operator $O_1(x)$ and those carrying the vertex corresponding
to the operator $O_2(x)$ in Eq. (3) (see \cite{haki3}).
The LO subprocesses contributing to the partonic cross section
are given by
\begin{eqnarray}
g + g \rightarrow g + {\rm A}\,, \qquad q + \bar q \rightarrow g + {\rm A}\,,
 \qquad q(\bar q) + g \rightarrow q(\bar q) + {\rm A}\,.
\end{eqnarray}
The matrix elements squared do not differ from those derived for the scalar
${\rm H}$ provided $n=4$, see \cite{spdj}, \cite{kasc}, which implies
that the LO double differential partonic cross sections 
are the same for both bosons except for an overall constant given by 
$F_{\rm B}(\tau)$ in Eq. (5).
In NLO one has to compute the one-loop virtual corrections to the processes
in Eq. (13) above and to add the contributions from the following 
two-to-three-body reactions
\begin{eqnarray}
&& g+ g \rightarrow g + g + {\rm A}\,, \qquad 
g + g \rightarrow q_i + \bar {q}_i + {\rm A}\,,
\\[2ex]
&& q+ \bar q \rightarrow g + g + {\rm A}\,,\qquad 
q_1 + \bar {q}_2 \rightarrow q_1 + \bar {q}_2 + {\rm A} \quad q_1 \not = q_2\,,
\nonumber\\[2ex]
&& q + \bar q \rightarrow q_i + \bar {q}_i + {\rm A}\quad q_i \not = q\,,
\qquad q + \bar q \rightarrow q + \bar q + {\rm A}\,,
\\[2ex]
&& q_1 + q_2 \rightarrow q_1 + q_2 + {\rm A} \quad q_1 \not = q_2\,, \qquad
q+ q \rightarrow q + q +{\rm A}\,,
\\[2ex]
&& q(\bar q) + g \rightarrow q(\bar q) + g + {\rm A}\,.
\end{eqnarray}
After renormalization of the strong coupling constant $\alpha_s$ and
mass factorization which are carried out in the ${\overline {\rm MS}}$-scheme
we obtain the NLO corrected coefficient functions according to the
procedure in \cite{rasm}. The coefficient functions are as long as in the case
of H-production so that they cannot be explicitly presented.
However the differences between the results for the H and the 
A are so small that we can show them below. If we put
for simplicity $G_{\rm H}=G_{\rm A}=G$ and $m_{\rm H}=m_{\rm A}=m$
the differences between 
the soft-plus-virtual differential cross sections are given by
\begin{eqnarray}
&&s^2 \frac{d^2~\sigma^{\rm S+V}_{gg\rightarrow g~{\rm A}}}{d~td~u}
-s^2 \frac{d^2~\sigma^{\rm S+V}_{gg\rightarrow g~{\rm H}}}{d~td~u}=
\nonumber\\[2ex]
&&\pi\,\delta(s+t+u-m^2)\,G^2\,\left (\frac{\alpha_s(\mu^2)}{4\pi}\right )^2\,
\frac{N}{(N^2-1)^2}\Bigg [2\,|M^{(1)}_{gg\rightarrow g~{\rm B}}|^2 \Bigg ]\,,
\\[2ex]
&&s^2 \frac{d^2~\sigma^{\rm S+V}_{q\bar q\rightarrow g~{\rm A}}}{d~td~u}
-s^2 \frac{d^2~\sigma^{\rm S+V}_{q \bar q\rightarrow g~{\rm H}}}{d~td~u}=
\nonumber\\[2ex]
&&\pi\,\delta(s+t+u-m^2)\,G^2\,\left (\frac{\alpha_s(\mu^2)}{4\pi}\right )^2\,
\frac{1}{N^2}\Bigg [2\,C_A\,|M^{(1)}_{q \bar q\rightarrow g~{\rm B}}|^2
\nonumber\\[2ex]
&&+\Bigg (C_F-C_A\Bigg )|MB^{(1)}_{q\bar q\rightarrow g~{\rm B}}|^2\Bigg ]\,,
\\[2ex]
&&s^2 \frac{d^2~\sigma^{\rm S+V}_{qg\rightarrow q~{\rm A}}}{d~td~u}
-s^2 \frac{d^2~\sigma^{\rm S+V}_{qg\rightarrow q~{\rm H}}}{d~td~u}=
\nonumber\\[2ex]
&&\pi\,\delta(s+t+u-m^2)\,G^2\,\left (\frac{\alpha_s(\mu^2)}{4\pi}\right )^2\,
\frac{1}{N(N^2-1)}\Bigg [2\,C_A\,|M^{(1)}_{qg\rightarrow q~{\rm B}}|^2
\nonumber\\[2ex]
&&+\Bigg (C_F-C_A\Bigg )|MB^{(1)}_{qg\rightarrow q~{\rm B}}|^2 \Bigg ]\,.
\end{eqnarray}
where we have added to the righthand side of Eqs. (19) and (20) the contributions
coming from the interference of the graphs in Figs. 1b, 1d with those in 
Figs. 2b, 2d which are shown in \cite{haki3}.
The colour factors of the group $SU(N)$ are given 
by $C_A=N$ and $C_F=(N^2-1)/(2N)$
and the Born matrix elements squared belonging to the processes in Eq. (13)
are equal to
\begin{eqnarray}
|M^{(1)}_{gg\rightarrow g~{\rm B}}|^2&=&N(N^2-1)\,\frac{1}{stu}\,\Bigg [
s^4+t^4+u^4+m^8\Bigg ]\,,
\\[2ex]
|M^{(1)}_{q\bar q\rightarrow g~{\rm B}}|^2&=&C_A\,C_F\,\frac{1}{s}\,
\Bigg [t^2+u^2\Bigg ]\,,
\\[2ex]
 |M^{(1)}_{qg\rightarrow q~{\rm B}}|^2&=&C_A\,C_F\,\,\frac{1}{u}\,
\Bigg [-s^2-t^2\Bigg ]\,.
\end{eqnarray}
The differences above can be wholly attributed to the virtual corrections 
and not to the soft gluon contributions which are the same for both H
and A-production. These virtual corrections also entail some extra 
terms denoted by
\begin{eqnarray}
|MB^{(1)}_{gg\rightarrow g~{\rm B}}|^2
&=&\frac{2}{3}\,N\,(N^2-1)\,\frac{m^2}{s\,t\,u}\,
\Big [s\,t\,u+m^2\, \Big (s\,t+s\,u+t\,u\Big )\Big ]\,,
\\[2ex]
|MB^{(1)}_{q\bar q\rightarrow g~{\rm B}}|^2&=&C_A\,C_F\,\Big (-t-u\Big )\,,
\\[2ex]
|MB^{(1)}_{qg\rightarrow q~{\rm B}}|^2&=&C_A\,C_F\,\Big (s+t\Big )\,.
\end{eqnarray}
Denoting the two-to-three-body reactions by 
\begin{eqnarray}
a(p_1)+b(p_2) \rightarrow c(-p_3)+d(-p_4)+{\rm A}(-p5)\,, 
\qquad s_4=(p_3+p_4)^2\,,
\end{eqnarray}
then the differences between the partonic cross sections due to the 
subprocesses in Eqs. (14)-(17) are equal to
\begin{eqnarray}
&& s^2 \frac{d^2~\sigma^{\rm HARD}_{gg\rightarrow gg~{\rm A}}}{d~t~d~u}
-s^2 \frac{d^2~\sigma^{\rm HARD}_{gg\rightarrow gg~{\rm H}}}{d~t~d~u}=
\nonumber\\[2ex]
&&\pi\,G^2\,\left (\frac{\alpha_s(\mu^2)}{4\pi}\right )^2\,\frac{N^2}{N^2-1}\,
\Bigg [-4\,\ln \frac{t u-m^2 s_4}{(s_4-t)(s_4-u) }-\frac{17}{3}\Bigg ]\,,
\\[2ex]
&& s^2 \frac{d^2~\sigma^{\rm HARD}_{gg\rightarrow q\bar q~{\rm A}}}{d~t~d~u}
-s^2 \frac{d^2~\sigma^{\rm HARD}_{gg\rightarrow q\bar q~{\rm H}}}{d~t~d~u}=
\nonumber\\[2ex]
&&\pi\,G^2\,\left (\frac{\alpha_s(\mu^2)}{4\pi}\right )^2\,\frac{n_f}{N^2-1}\,
\Bigg [C_A\Bigg \{\frac{2}{3}\Bigg \}+C_F\Bigg \{
2\,\ln \frac{t u-m^2 s_4}{(s_4-t)(s_4-u) } + 2\Bigg \}\Bigg ]\,,
\\[2ex]
&& s^2 \frac{d^2~\sigma^{\rm HARD}_{q\bar q\rightarrow gg~{\rm A}}}{d~t~d~u}
-s^2 \frac{d^2~\sigma^{\rm HARD}_{q\bar q\rightarrow gg~{\rm H}}}{d~t~d~u}=
\nonumber\\[2ex]
&&\pi\,G^2\,\left (\frac{\alpha_s(\mu^2)}{4\pi}\right )^2\,\frac{C_A~C_F}{N^2}\,
\Bigg [C_A\Bigg \{\frac{2}{3}\Bigg \}+C_F\Bigg \{
2\,\ln \frac{t u-m^2 s_4}{(s_4-t)(s_4-u) } + 2\Bigg \}\Bigg ]\,,
\\[2ex]
&& s^2 \frac{d^2~\sigma^{\rm HARD}_{q_1\bar q_2\rightarrow q_1\bar q_2~{\rm A}}}
{d~t~d~u}-s^2 \frac{d^2~\sigma^{\rm HARD}_{q_1\bar q_2 \rightarrow q_1\bar q_2~
{\rm H}}}{d~t~d~u}=
\nonumber\\[2ex]
&&\pi\,G^2\,\left (\frac{\alpha_s(\mu^2)}{4\pi}\right )^2\,\frac{C_A~C_F}{N^2}\,
\Bigg [-2\,\ln \frac{t u-m^2 s_4}{(s_4-t)(s_4-u) } - 1\Bigg ]\,,
\\[2ex]
&& s^2 \frac{d^2~\sigma^{\rm HARD}_{q_1\bar q_1\rightarrow q_i\bar q_i~{\rm A}}}
{d~t~d~u}-s^2 \frac{d^2~\sigma^{\rm HARD}_{q_1\bar q_1 \rightarrow q_i\bar q_i
~{\rm H}}}{d~t~d~u}=
\nonumber\\[2ex]
&&\pi\,G^2\,\left (\frac{\alpha_s(\mu^2)}{4\pi}\right )^2\,
\frac{(n_f-1)~C_A~C_F}{N^2}\, \Bigg [-\frac{2}{3}\Bigg ]\,,
\\[2ex]
&& s^2 \frac{d^2~\sigma^{\rm HARD}_{q\bar q\rightarrow 
q\bar q~{\rm A}}}{d~t~d~u}
-s^2 \frac{d^2~\sigma^{\rm HARD}_{q\bar q \rightarrow 
q\bar q~{\rm H}}}{d~t~d~u}=
\nonumber\\[2ex]
&&\pi\,G^2\,\left (\frac{\alpha_s(\mu^2)}{4\pi}\right )^2\,
\frac{C_F}{N^2}\, \Bigg [C_A\Bigg \{ 
-2\,\ln \frac{t u-m^2 s_4}{(s_4-t)(s_4-u) } - \frac{5}{3}\Bigg \}
\nonumber\\[2ex]
&&+\frac{s~s_4~((s-m^2)^2+s_4^2-2~t~u)}{8(s_4-t)^2(s_4-u)^2}
+\frac{(s-m^2)^2+s_4^2-2~t~u}{4(s_4-t)(s_4-u)}
\nonumber\\[2ex]
&&+\frac{(s-m^2)^2+s_4^2-2~t~u+6s s_4}{4ss_4}
\ln \frac{t u-m^2 s_4}{(s_4-t)(s_4-u)}+\frac{9}{4}\Bigg ]\,,
\\[2ex]
&& s^2 \frac{d^2~\sigma^{\rm HARD}_{q_1q_2\rightarrow q_1q_2~{\rm A}}}{d~t~d~u}
-s^2 \frac{d^2~\sigma^{\rm HARD}_{q_1q_2 \rightarrow q_1q_2~{\rm H}}}{d~t~d~u}=
\nonumber\\[2ex]
&&\pi\,G^2\,\left (\frac{\alpha_s(\mu^2)}{4\pi}\right )^2\,\frac{C_A~C_F}{N^2}\,
\Bigg [-2\,\ln \frac{t u-m^2 s_4}{(s_4-t)(s_4-u) } - 1\Bigg ]\,,
\\[2ex]
&& s^2 \frac{d^2~\sigma^{\rm HARD}_{qq\rightarrow qq~{\rm A}}}{d~t~d~u}
-s^2 \frac{d^2~\sigma^{\rm HARD}_{qq \rightarrow qq~{\rm H}}}{d~t~d~u}=
\nonumber\\[2ex]
&&\pi\,G^2\,\left (\frac{\alpha_s(\mu^2)}{4\pi}\right )^2\,\frac{C_F}{N^2}\,
\Bigg [C_A\Bigg \{-2\,\ln \frac{t u-m^2 s_4}{(s_4-t)(s_4-u) } - 1\Bigg \}
\nonumber\\[2ex]
&&+\frac{s^2+s_4^2}{4~(s_4-t)(s_4-u)} \ln \frac{s~s_4}{t u-m^2 s_4} 
-\frac{3}{2}\,\ln \frac{t u-m^2 s_4}{(s_4-t)(s_4-u)} \Bigg ]\,,
\\[2ex]
&& s^2 \frac{d^2~\sigma^{\rm HARD}_{qg\rightarrow qg~{\rm A}}}{d~t~d~u}
-s^2 \frac{d^2~\sigma^{\rm HARD}_{qg \rightarrow qg~{\rm H}}}{d~t~d~u}=
\nonumber\\[2ex]
&&\pi\,G^2\,\left (\frac{\alpha_s(\mu^2)}{4\pi}\right )^2\,\frac{1}{N}\,
\Bigg [C_A\Bigg \{-2\,\ln \frac{t u-m^2 s_4}{(s_4-t)(s_4-u) } - 1\Bigg \}
\nonumber\\[2ex]
&& +C_F\Bigg \{ \ln \frac{t u-m^2 s_4}{(s_4-t)(s_4-u) } - \frac{1}{2}\Bigg \}
\Bigg ]\,,
\end{eqnarray}
where the meaning of the superscript ${\rm HARD}$ is explained in \cite{rasm}. 
From these expressions we infer that  
the partonic cross sections (coefficient functions) for ${\rm H}$
and ${\rm A}$ are equal in LO and almost equal in NLO. This means that apart
from the overall normalisation due to the constant $G_{\rm B}$ there will 
not be any difference in the shapes of the double differential cross sections.
We show this in Fig. 1 where we plot the ratio
\begin{eqnarray}
R=\frac{d\sigma_{\rm A}}{d\sigma_{\rm H}}\,,
\end{eqnarray}
for $d\sigma_{\rm B}=d\sigma_{\rm B}/dp_T$ and proton-proton collisions 
at the LHC 
with $\sqrt{S}=14~{\rm TeV}$. For these and the next plots we have
adopted the parton density set MRST98 (LO,lo05a.dat) 
\cite{mrst98} for the LO calculations with
$\Lambda_5^{\rm NLO}=130.5~{\rm MeV}$ as input for the leading log running
coupling constant. For the NLO cross sections we have chosen the set
MTST99 (NLO,cor01.dat) \cite{mrst99} with $\Lambda_5^{\rm NLO}=220~{\rm MeV}$ 
as input for the next-to-leading log running coupling constant. Furthermore the 
factorization/renormalization scale is chosen to be
$\mu^2=\mu_r^2=p_T^2+M_{\rm B}^2$. For the
masses of the Higgs bosons we take $m_{\rm H}=m_{\rm A}=120~{\rm GeV/c^2}$
and the top quark mass is set equal to $m_t=173.4~{\rm GeV/c^2}$. Further
we have put $\tan \beta=1$. In the case of an infinite top quark mass,
(here we choose $m_t=173.4\times 10^3~{\rm GeV/c^2}$), 
we get $R^{\rm LO}=9/4$ irrespective of the values of $m_{\rm H}$ and
$m_{\rm A}$. This follows from Eq. (6) and the fact that the LO partonic
cross sections are the same for ${\rm H}$-production and ${\rm A}$-production. 
A finite $m_t$ as given above introduces a small effect and one 
gets $R^{\rm LO}=2.31$ which amounts to a shift
upwards of $0.06$ (see Fig. 1). 
In NLO the partonic cross sections differ a little bit
and ${\cal C}^2_{\rm H}=[1+22 \alpha_s/(4\pi)]~{\cal C}^2_{\rm A}$ (see
Eqs. (7),(8)). Therefore we expect a deviation from the $R^{\rm LO}$ result 
when $m_t$ is taken infinite in both the LO and NLO reactions.
However it turns out that both differences compensate each other.
The NLO corrected partonic cross section for A 
is larger than the one for H and one obtains
an upward shift $\Delta R^{\rm NLO}=0.26$. The shift due to the coefficient
function in Eq. (7) is negative and amounts to $\Delta R^{\rm NLO}=-0.24$.
Hence the actual value becomes $R^{\rm NLO}=2.27$ (see Fig. 1) which is very 
close to $R^{\rm NLO}=9/4$. If $m_t$ is finite one gets
again an upward shift of $0.06$ like in LO and $R^{\rm NLO}=2.33$ (see Fig. 1).
One can make similar plots for the rapidity $y$ distributions which yield the
same ratios as shown in Fig. 1 for the $p_T$ distributions.
The most important feature is that the ratios are independent of $p_T$ and $y$
showing the shape independence of the distributions on the parity 
of the Higgs boson (scalar versus pseudo-scalar). This behaviour was 
discovered for both the (pseudo)-scalar $p_T$ distributions and for the opening angle distribution between the (pseudo)-scalar bosons 
and the highest $p_T$-jet in the 
reaction $p +p \rightarrow ({\rm H\, or \, A}) + {\rm jet} +{\rm jet} + 'X'$ 
in \cite{field}.
From Fig. 1 and the observations made above it is clear that the 
ratios between the NLO and LO corrected cross sections (K-factors) are the same
for H-production  and A-production. This also holds for the variation of the 
NLO cross sections with respect to the mass factorization/renormalization
scales. They are given for H-production in \cite{flgra}, 
\cite{rasm} and we do not have to show them again for A-production. 
In Fig. 2 we present the $p_T$ distributions in NLO
for ${\rm A}$-production in 
proton-antiproton collisions at $\sqrt{S}=2~{\rm TeV}$ 
(Fermilab Tevatron, Run II) and in proton-proton collisions 
at $\sqrt{S}=14~{\rm TeV}$ (LHC).  Further we have chosen
$m_{\rm A}=91.9~{\rm GeV/c^2}$ and $\tan \beta=0.5$. The parton density set 
and the factorization scale are given above. From Fig. 2 we infer 
that the $p_T$-distributions decrease rather slowly as $p_T$ increases
and that the differential cross section for the Tevatron is two 
orders of magnitude smaller than the one predicted for the LHC. 
The latter observation also holds for the corresponding rapidity distributions 
shown in Fig. 3. They are obtained by integrating
$d^2\sigma_{\rm A}/(dp_T~dy)$ over the range
$p_{T,{\rm min}}<p_T<8~ p_{T,{\rm min}}$ with $p_{T,{\rm min}}=30~{\rm GeV/c}$.
The cross section for $p_T>8~ p_{T,{\rm min}}$ is negligible. Notice that
the range of the rapidity for A-production at the Tevatron is rather small. 
Finally we want to comment on the relative importance of the partonic 
subprocesses contributing to the hadronic differential cross section 
in Eq. (12). For the LHC ($\sqrt{S}=14~{\rm TeV}$) the $gg$-channel 
dominates and the $qg$-subprocess  
contributes about one third of the cross section. This is because
at these high energies the $x$-values of the gluon density $f_g^{\rm P}(x)$
is so small that it becomes much larger than the quark densities.
At lower energies like $\sqrt{S}=2~{\rm TeV}$ (Tevatron) the $x$-values
are larger so that the valence quark densities also play a role.
This explains why the contribution of the $qg$-subprocess is of the same 
magnitude as the one from the $gg$-channel for A-production at the Tevatron. 
\\
{\bf Acknowledgments.} We are indebted to V. Ravindran for discussions about
the relevance of the operator $O_2(x)$.
B. Field would like to thank S. Dawson and W. Kilgore
for discussions on the pseudo-scalar coupling. 

\centerline{\bf \large{Figure Captions}}
\begin{description}
\item[Fig. 1.]
The ratio $R$ in Eq. (37) plotted as a function of $p_T$ for
$\sqrt{S}=14~{\rm TeV}$ and $\mu^2=p_T^2+m_{\rm B}^2$ with $m_{\rm H}=m_{\rm A}
=120~{\rm GeV/c^2}$; $R^{\rm LO}(m_t=\infty)$ (dotted line),
$R^{\rm LO}(m_t=173.4)$ (solid line) $R^{\rm NLO}(m_t=\infty)$ (dot-dashed
line) $R^{\rm NLO}(m_t=173.4)$ (dashed line).
\item[Fig. 2.]
The transverse momentum distribution $d\sigma_{\rm A}/dp_T$ 
with $\mu^2=p_T^2+m_{\rm A}^2$, $m_{\rm A}=91.9~{\rm GeV/c^2}$, 
$\tan \beta=0.5$;
$\sqrt{S}=14~{\rm TeV}$ (solid line), $\sqrt{S}=2~{\rm TeV}$ (dashed line).
\item[Fig. 3.]
The rapidity distribution $d\sigma_{\rm A}/dy$  calculated from 
the integral of $d^2\sigma_{\rm A}/(dp_T~dy)$  
between $8~p_{T,{\rm min}}>p_T>p_{T,{\rm min}}$
and $p_{T,{\rm min}}=30~{\rm GeV/c}$.
Input parameters are $\mu^2=p_{T,{\rm min}}^2+m_{\rm A}^2$, 
$m_{\rm A}=91.9~{\rm GeV/c^2}$, $\tan \beta=0.5$;
$\sqrt{S}=14~{\rm TeV}$ (solid line), $\sqrt{S}=2~{\rm TeV}$ (dashed line).
\end{description}
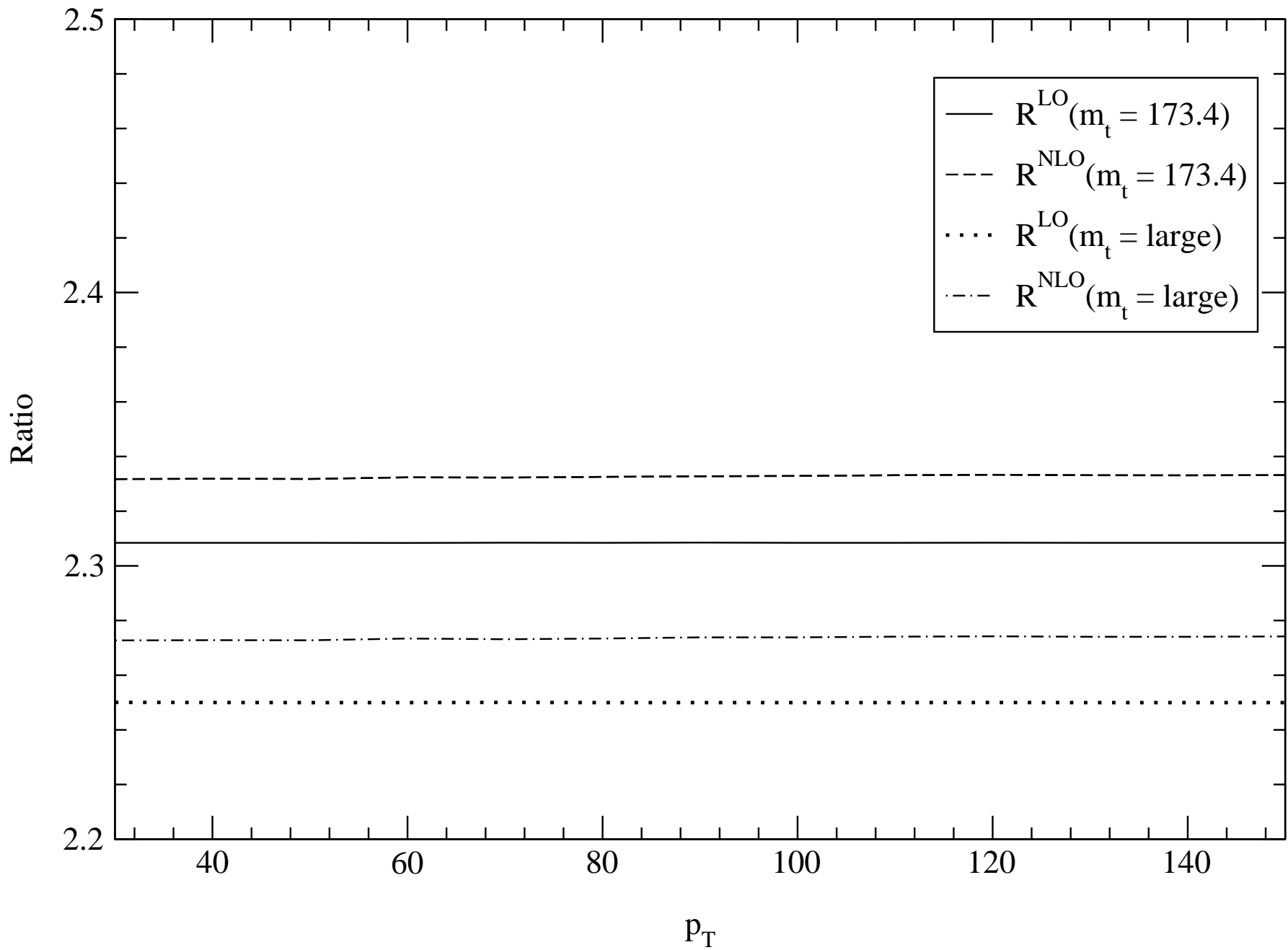
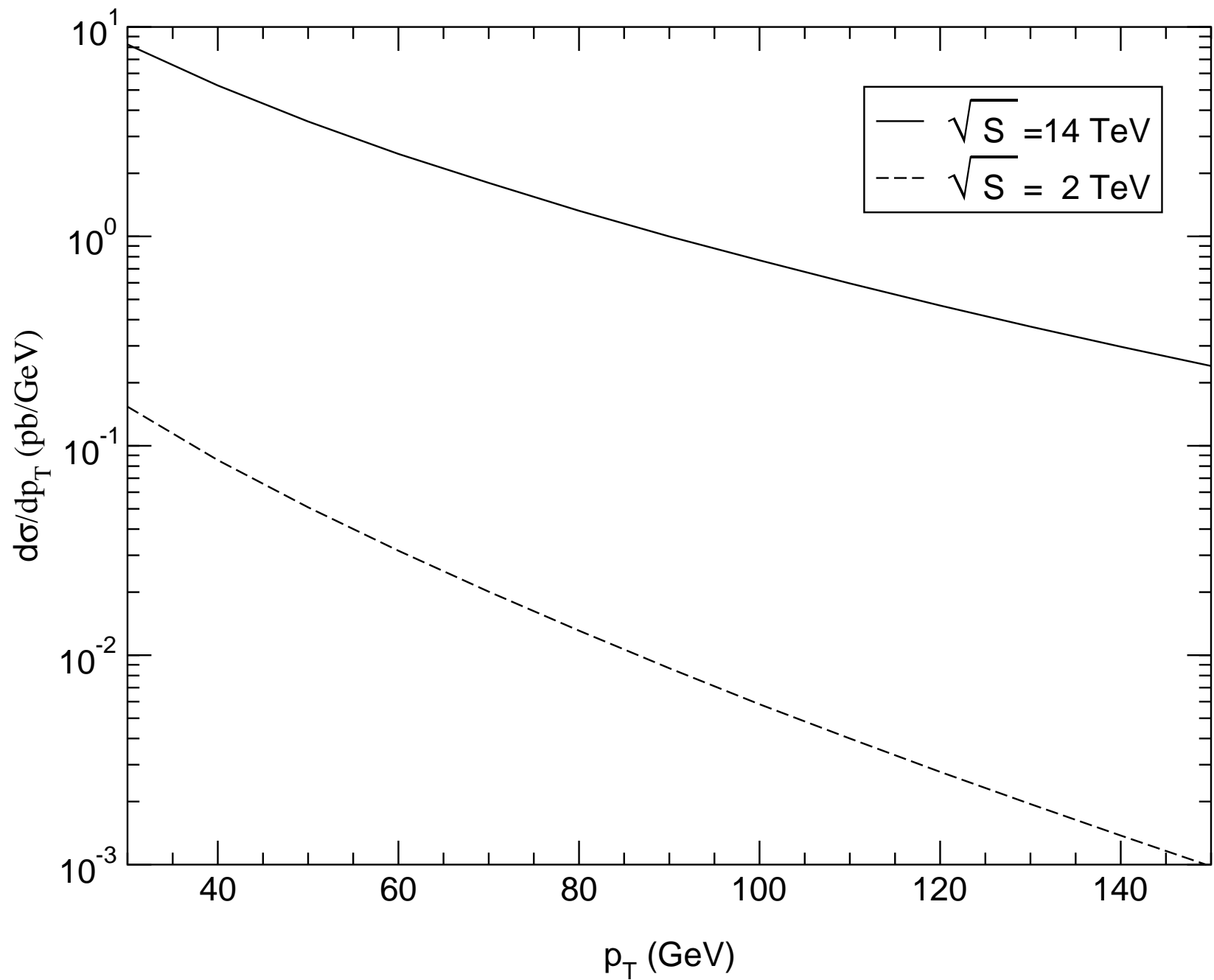
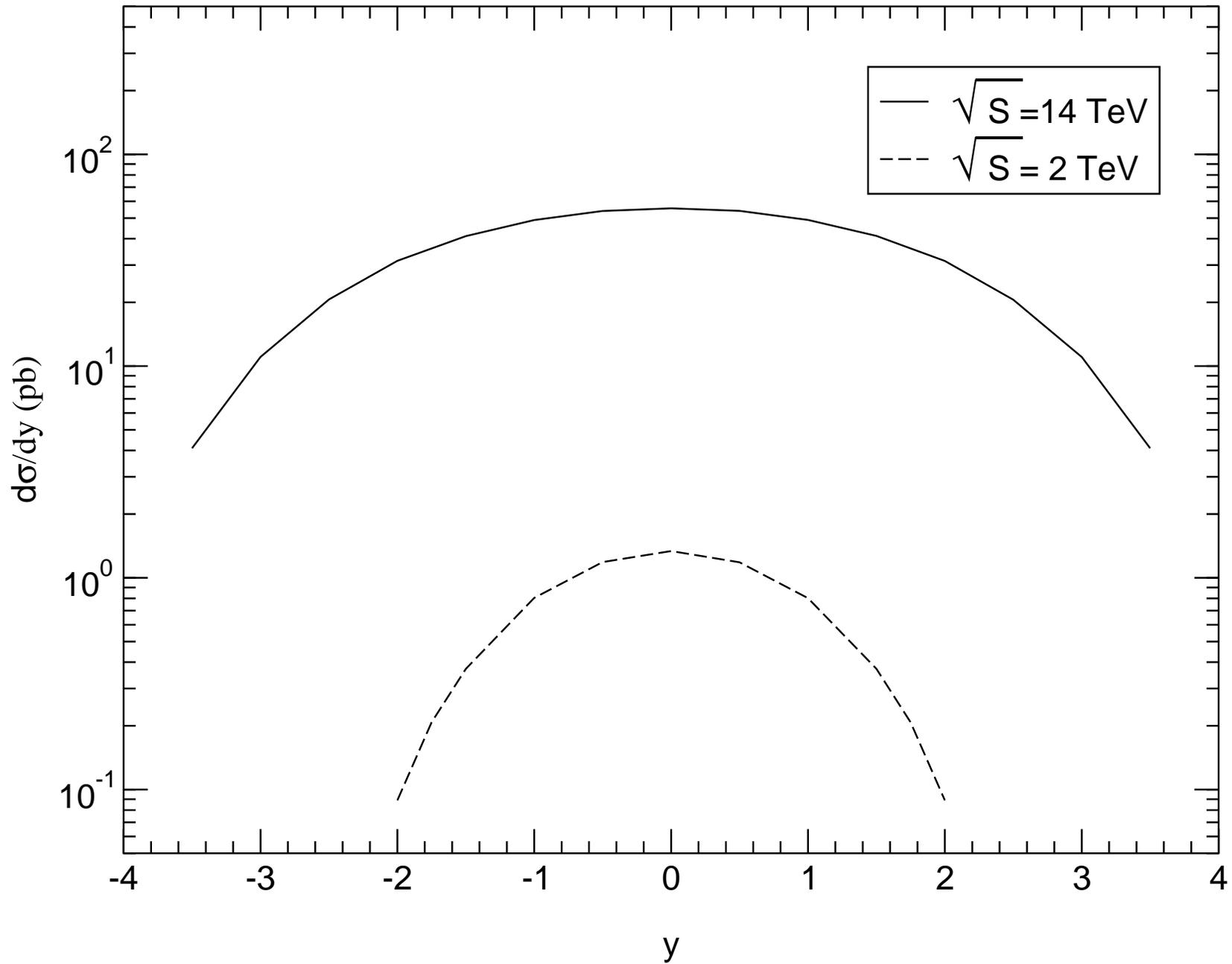
\end{document}